\documentclass{iau} 
\usepackage{graphicx}

\title[Tidally-disrupted Molecular Clouds] 
{Tidally-disrupted Molecular Clouds falling to the Galactic Center}

\author[M. Tsuboi et al.]   
{Masato Tsuboi$^1$, Yoshimi Kitamura$^1$, Kenta Uehara$^2$, \\Ryosuke Miyawaki$^3$, 
 \and Atsushi Miyazaki$^4$}

\affiliation{$^1$Institute of Space and Astronautical Science, JAXA,
 Sagamihara, Kanagawa 252-5210, Japan\\ email: {\tt tsuboi@vsop.isas.jaxa.jp} \\[\affilskip]
$^2$Department of Astronomy, The University of Tokyo, Bunkyo, Tokyo 113-0033, Japan\\[\affilskip]
$^3$College of Arts and Sciences, J.F. Oberlin University, Machida, Tokyo 194-0294, Japan\\[\affilskip]$^4$
Japan Space Forum, Kandasurugadai, Chiyoda-ku, Tokyo 101-0062, Japan}

\pubyear{2016}
\volume{xxx}  
\jname{The Multi-Messenger Astrophysics of the Galactic Centre}
\editors{Roland Crocker, eds.}
\begin{document}

\maketitle

\begin{abstract}
We found a molecular cloud connecting from the outer region to the ``Galactic Center Mini-spiral (GCMS)" which is a bundle of the ionized gas streams adjacent to Sgr A$^\ast$.   
The molecular cloud has a filamentary appearance which is prominent in the CS $J=2-1$ emission line and is continuously connected with the GCMS.  The velocity of the molecular cloud is also continuously connected with that of the ionized gas in the GCMS observed in the H42$\alpha$ recombination line. The morphological and kinematic  relations suggest that the molecular cloud is falling from the outer region to  the vicinity of Sgr A$^\ast$, being disrupted by the tidal shear of Sgr A$^\ast$ and ionized by UV emission from the Central Cluster.
We also found the SiO $J=2-1$ emission in the boundary area between the filamentary molecular cloud and the GCMS. There seems to exist shocked gas in the boundary area.  
\end{abstract}

\firstsection 
\section{Introduction}
The Galactic Center is  the nuclear region of the nearest spiral galaxy, Milky Way. The environment is unique in the galaxy because the region contains several peculiar objects. First, Sagittarius A$^\ast$ (Sgr A$^\ast$) is a counter part of the Galactic Center Black Hole (GCBH) in the regime from radio to X-ray, which  is located very near the dynamical center of the galaxy (e.g. \cite[Reid \etal\ 2003]{Reid}) and has a mass of $\sim4\times10^6 $M$_\odot$ (e.g. \cite[Ghez \etal\ 2008]{Ghez}; \cite[Gillessen \etal\ 2009]{Gillessen}).  Second, the ``Circum-Nuclear Disk (CND)" is a torus-like molecular gas rotating around Sgr A$^\ast$, which is identified at the distance to a few pc (e.g. \cite[G\"usten \etal\ 1987]{Guesten}).   A bundle of the ionized gas streams is located in the inner cavity of the CND. This is called ``Galactic Center Mini-spiral (GCMS)"  (e.g. \cite[Ekers \etal\ 1983]{Ekers1983}; \cite[Lo\&Claussen 1983]{LO1983}). The stretched appearance and kinematics with large velocity gradient suggest that it is a tentative structure surrounding Sgr A$^\ast$. Finally, the Central cluster is a star cluster centered at Sgr A$^\ast$.  Though the cluster concentrates within $r < 0.5$ pc, it contains $\sim$100 OB and WR stars  (e.g. \cite[Genzel \etal\ 1996]{Genzel1996}; \cite[Paumard \etal\ 2006]{Paumard}). There remains considerable controversy about the cluster origin. 
It would be difficult to form the cluster in a way by which stars are usually formed in the galactic disk because of the following reasons. The tidal force of Sgr A$^\ast$ must have a serious effect on the star formation because the minimum number H$_2$ density for stabilization toward the tidally shearing is $n($H$_2)>3\times10^8$ cm$^{-3}$ at $r < 0.5$ pc (\cite[Christopher \etal\ 2005]{Christopher}, \cite[Montero-Casta\~{n}o \etal\ 2009]{Montero}, \cite[Tsuboi \etal\ 2011]{Tsuboi2011}). In addition, the strong Lyman continuum radiation from the early type stars in the cluster ionizes rapidly the ISM in the region. 
Because the star formation in the vicinity of the Sgr A$^\ast$ requires an additional mechanism to overcome the difficulties, it is still an open question how the Central cluster is formed.
Two distinct but compatible scenarios for the formation of the Central cluster have been proposed. One is current in-situ star formation in such extreme environment of the vicinity of Sgr A$^\ast$. Previous observations have revealed that there are many  molecular clumps in the CND(e.g. \cite[Montero-Casta\~{n}o \etal\ 2009]{Montero}, \cite[Mart\'{\i}n \etal\ 2012]{Martin}). Cloud-cloud collision in the CND is proposed as the star formation  mechanism in the region (e.g. \cite[Jalali \etal\ 2014]{Jalali}). Unfortunately, we find no cradle dense molecular clouds around massive stars in the inner cavity of the CND at present. The other is that the molecular cloud is falling from the region considerably far from Sgr A$^\ast$ to the vicinity after star formation have already started in the cloud. Thus we searched the falling molecular cloud with star formation in the Sgr A region. 

\begin{figure}[b]
\begin{center}
 \includegraphics[width=5.5in]{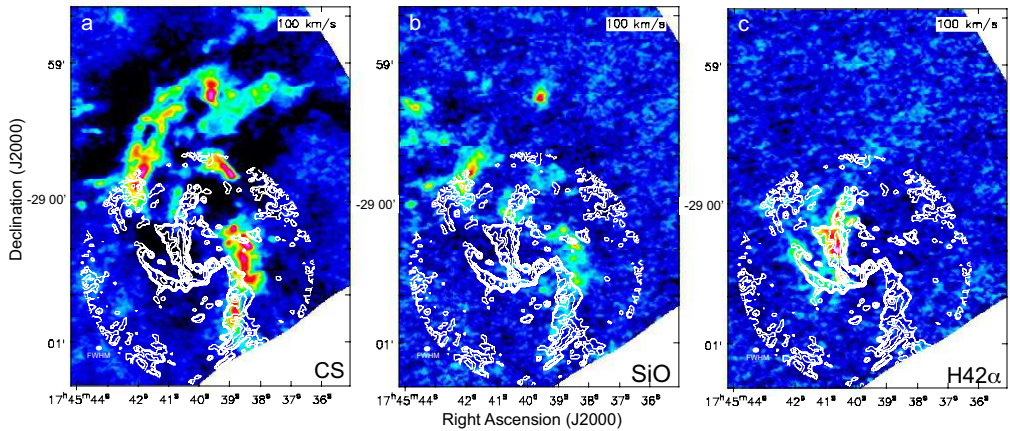} 
 \caption{Integrated  intensity maps of the Sgr A region:  {\bf a}: in the CS $J=2-1$ emission line,  {\bf b}:  in the  SiO $v=0, J=2-1$ emission line,  {\bf c}: in the  H42$\alpha$ recombination line. The central velocity is $V_{\mathrm{LSR}}=100$ km s$^{-1}$. The angular resolution is  shown on the lower-left corner of the each panel as a filed oval. The integration velocity widths are 10 km s$^{-1}$ in {\bf a} and {\bf b} and 20 km s$^{-1}$ in {\bf c}. The contours show the continuum emission of the ``Galactic Center Mini-spiral" at 100 GHz for comparison (\cite[Tsuboi \etal\ 2016a]{Tsuboi2016a}). }
   \label{fig1}
\end{center}
\end{figure}

\section{Observation and Data Reduction}
We have produced a 137 pointing mosaic of the 12-m array and a 52 pointing mosaic of the 7-m array (ACA) which cover a $330" \times 330"$ area including  the ``Galactic Center 50 km s$^{-1}$ Molecular Cloud (50MC)" and the CND in the CS $J=2-1$($97.980953$ GHz), and SiO $v=0~J=2-1$($86.846995$ GHz) emission lines by ALMA Cy.1 (2012.1.00080.S, PI Tsuboi, M.). The frequency range also includes the H42$\alpha$ recombination line($85.6884$ GHz). 
We have detected these emission lines from both the two regions. 
The data have an angular resolution of $(2.2-2.5)" \times (1.5-1.8)"$ using ``natural weighting" in UV sampling. The ALMA synthesized beam is approximately 4 times smaller than those of previous molecular line observations (e.g. \cite[Montero-Casta\~{n}o \etal\ 2009]{Montero}, \cite[Mart\'{\i}n \etal\ 2012]{Martin}).  
J0006-0623, J1517-2422, J717-3342,  J1733-1304, J1743-3058, J1744-3116 and J2148+0657 were used as phase calibrators. 
The flux density scale was determined using Titan, Neptune and Mars. The calibration and imaging of the data were done by CASA (\cite[McMullin \etal\  2007]{McMullin}). The continuum emission of  the GCMS and Sgr A$^\ast$  was subtracted from the spectral data using the CASA task UVCONTSUB. Because the observation has a large time span of one year and seven months, the flux uncertainty is as large as 10 \%. 

\section{Results and  Discussion}
Figures 1a and 1b show the integrated intensity maps  (pseudo color) of the CND in the CS $J=2-1$ and SiO $v=0, J=2-1$ emission lines, respectively. They are cut from the large area mosaic data mentioned in the previous section. The central velocity is $V_{\mathrm{LSR}}=100$ km s$^{-1}$ both in the maps. The velocity width of each panel is 10 kms$^{-1}$. The CS emission line is a dense molecular gas  tracer with $n$(H$_2$)$\gtrsim 10^4$ cm$^{-1}$, while the SiO emission line is a tracer of shocked molecular gas with shock velocity of $V_{\mathrm{s}}\gtrsim 50$km s$^{-1}$.
The angular resolutions are $2.3" \times 1.6"$ for the CS $J=2-1$ emission line and $2.5" \times 1.8"$ for the SiO $v=0, J=2-1$ emission line, respectively. They are shown on the lower-left  corners of the two panels as filed ovals.
The contours in the figures show the continuum emission of the GCMS at 100 GHz for comparison (\cite[Tsuboi \etal\ 2016a]{Tsuboi2016a}). 

We found a filamentary molecular cloud extending from $\alpha\sim17^h45^m43^s$,  $\delta\sim-29^\circ00'00"$ to $\alpha\sim17^h45^m37^s$,  $\delta\sim-28^\circ58'40"$ in figure 1a. The north end of the molecular cloud reaches beyond the CND; $r\gtrsim 6$ pc. This molecular cloud is prominent in the velocity ranges from $V_{\mathrm{LSR}}=+70$ km s$^{-1}$ to $V_{\mathrm{LSR}}=+120$ km s$^{-1}$. Along the velocity increases as approaching the south end.  
Although the filamentary molecular cloud is prominent in the CS 2-1 emission line, this had never been identified by existing telescopes because it is deeply embedded in the CND. However, this is distinguishable clearly from other clouds belong to the CND with the curved filamentary appearance revealed by the high angular resolution of ALMA.  
On the other hand, the filamentary molecular cloud is not prominent  in the SiO emission line except around the south end, $\alpha\sim17^h45^m42^s$,  $\delta\sim-28^\circ59'50"$. This indicates that the cloud suffers from strong shock around  the south end although there is no evidence of strong shock in the other parts of the cloud. 
The south end of the molecular cloud has the very wide velocity width of $\Delta V\gtrsim70$ km s$^{-1}$. 
While there is another component along the ``Western arc" of the GCMS in figures 1a and 1b.  This component probably belongs to the CND. Meanwhile, figure 1c  shows the integrated  intensity map  (pseudo color)  of the GCMS in the H42$\alpha$ recombination line (pseudo color). The H42$\alpha$ recombination line is an ionized gas tracer. The south end seems to be connected continuously to the north extension of the ``Eastern arm" and ``Northeastern arm" of the GCMS (\cite[Tsuboi \etal\ 2016b]{Tsuboi2016b}).

Although the velocity width of  the north extension is as large as $\Delta V\sim100$ km s$^{-1}$, the central velocity is $V_{\mathrm{LSR}}\sim100$ km s$^{-1}$ (\cite[Tsuboi \etal\ 2016b]{Tsuboi2016b}) and overlaps to that of the south end of the molecular cloud.  The central velocity  of the molecular cloud is also continuous with that of the ionized gas in the arms.
The morphological and kinematic properties suggest that the molecular cloud continues to the ionized gas physically. As mentioned above, the shocked gas traced by the SiO emission is located around the boundary of these. The molecular cloud is ionized by the Lyman continuum from the Central cluster as approaching to Sgr A$^\ast$ and expands at several hundreds times because the gas temperature increases from several $10$ K to $1\times10^4$ K. Therefore the interaction between the molecular gas and ionized gas probably produces a shock wave around the interface. 

The filamentary molecular cloud is probably falling  from the outer region to the Sgr A region and being  disrupted by the tidal shear of Sgr A$^\ast$. 
The star formation in the cloud may have already started by an external trigger like cloud-cloud collision when the cloud was in the region far from Sgr A$^\ast$.  Shock waves induced by cloud-cloud collision may form massive molecular cloud cores (e.g. \cite[Inoue \& Fukui 2013] {Inoue}).  A possible remnant of such cloud-cloud collision is observed in the 50MC (\cite[Tsuboi \etal\ 2015]{Tsuboi2015}). In the case of nearly head-on collision, the molecular clouds lose their angular momentums and begin to fall toward Sgr A$^\ast$. Moreover, the formed cores in the clouds  can not be destroyed by the tidal force because of their high densities. 
The massive cores quickly grow to OB stars. The speculation is consistent with the fact that the OB stars are observed on the dust peaks  in  the GCMS (\cite[Tsuboi \etal\ 2016a]{Tsuboi2016a}). 
While less massive cores gradually grows to low mass stars. 
These may be still  on the stage of protostars in the GCMS. Recently, several  half-shell-like ionized components are found  near Sgr A$^\ast$ by JVLA, which would be low mass protostars (\cite[Yusef-Zadeh \etal\ 2015]{Yusef-Zadeh2015}). They are located exclusively on the approaching tip of the ``Northeastern arm". 
These suggest that the filamentary cloud and  GCMS play a role  in transferring material including protostars from the outer region of the Sgr A complex to the vicinity of Sgr A$^\ast$. Some of the falling stars may be captured around Sgr A$^\ast$ and others may be scattered far from Sgr A$^\ast$.

\end{document}